\documentclass[runningheads]{llncs}
\usepackage[T1]{fontenc}
\usepackage{graphicx}
\usepackage{amssymb}
\usepackage{amsmath}
\usepackage{todonotes}
\usepackage{comment}

\usepackage{tikz}
\usetikzlibrary{quantikz}

\newcommand{\cent}[0]{\mbox{\textcent}}
\newcommand{\dollar}{\$}
\newcommand{\mymatrix}[2]{\left( \begin{array}{#1} #2 \end{array} \right)}

\newcommand{\mypar}[1]{\left( #1 \right)}

\newcommand{\modp}{\mathtt{MOD_p}}

\usepackage{subcaption}

\begin{document}
\title{A Representative Framework for Implementing Quantum Finite Automata on Real Devices}
%
%
\author{Aliya Khadieva\inst{1,2}\orcidID{0000-0003-4125-2151} \and
\"Ozlem Saleh{\.i}\inst{3,4}\orcidID{0000-0003-2033-2881} \and
Abuzer Yakary{\i}lmaz\inst{1,4}\orcidID{0000-0002-2372-252X}}
\authorrunning{A. Khadieva et al.}
%
\institute{University of Latvia, R\={\i}ga, Latvia \and
Kazan Federal University, Kazan, Russia \and 
Institute of Theoretical and Applied Informatics, Polish Academy of Sciences, Gliwice, Poland \and
QWorld Association, Tallinn, Estonia, \url{https://qworld.net} \and 
Zavoisky Physical-Technical Institute, FRC Kazan Scientific Center of RAS 
\\
\email{aliya.khadi@gmail.com, ozlemsalehi@gmail.com, abuzer.yakaryilmaz@lu.lv}
}
\maketitle              
\begin{abstract}
 We present a framework for the implementation of quantum
finite automata algorithms designed for the language $ \modp = \{ a^{i\cdot p } \mid i \geq 0 \}$ on gate-based quantum computers. First, we compile the known theoretical
results from the literature to reduce the number of CNOT gates.
Second, we demonstrate techniques for modifying the algorithms based
on the basis gates of available quantum hardware in order to reduce circuit
depth. Lastly, we explore how the number of CNOT gates may be
reduced further if the topology of the qubits is known.
 
\end{abstract}
\vspace{-0.4cm}
\section{Introduction}
\label{sec:intro}
\vspace{-0.3cm}
Finite automata serves as a well established model for data stream processing. Over the past few decades, the quantum analogue known as quantum finite automaton (QFA) has been widely investigated. Quantum finite automata can be exponentially more memory (state) efficient compared to probabilistic finite automata (PFAs) \cite{af98}: For any prime number $p$, any PFA recognizing the language $ \modp = \{ a^{i\cdot p } \mid i \geq 0 \}$ requires at least $p$ states, while bounded-error QFAs use only $ O (\log p)$ states. Notably, concepts like quantum fingerprinting ~\cite{bcwd2001} and quantum hashing~\cite{aavz2016} can be considered as generalizations of the QFA algorithm for the $\modp$ language. Moreover, the idea of comparing hashes for two different objects modulo $p$, which lies at the core of recognizing the $\modp$ language has applications in branching programs \cite{kk2017,kkk2022,agky16} and online algorithms \cite{kk2019disj,kk2022}.
%

Although, the QFA design for $\modp$ is straightforward, its implementation on gate-based quantum computers faces certain difficulties, as highlighted in \cite{bsocy2021,kalis18,SalehiY2021}. Currently, we are in the noisy intermediate-scale quantum era \cite{preskill2018quantum}, and one of the main problems with the available hardware (including near-future devices) is the noise introduced by quantum operators and measurements. Note that any quantum program is transpiled into two-qubit and single-qubit gates before its execution on real machines. Compared to single-qubit gates, two-qubit gates are more costly to implement and susceptible to errors. The transpilation process works based on a pre-determined set of basis gates, and this set commonly has CNOT (controlled NOT) gate as the only two-qubit gate. Consequently, one of the key measures when assessing the complexity of QFA implementations is the number of CNOT gates. Moreover, the depth of the circuit is also another factor that effects the decoherence and shorter circuits are preferable. 
Finally, in each real backend, qubits are organized in a certain topology, which determines the pair of gates on which two-qubit gates like CNOT can be applied. Hence, additional SWAP gates, that are composed of CNOT gates become necessary when transpiling any quantum program.  
%
%

The basic step in implementing a fingerprinting algorithm involves
 rotating a target qubit by a set of $d$ distinct angles, a task which can be implemented by applying a uniformly controlled rotation gate \cite{af98}. Möttönen et al. propose an efficient implementation of such operator in \cite{mottonen2006decompositions}, which uses $O(d)$ CNOT gates and $O(d)$ single-qubit gates. However, this method requires rotation angles to be modified, resulting in remarkably smaller angles. Note that for the current quantum computers, there is also a limit to the precision of implementing such rotations \cite{maldonado2022error}. In response to those challenges, we propose a technique for implementing uniformly controlled rotation that enables us to obtain a  trade-off between the number of CNOT gates and the precision of rotation angles. The technique combines Möttönen's approach with a particular decomposition of rotation gates. 
 Namely, for any integer value of parameter $t$ such that $ 0<t<\log_2d-4$, the CNOT-cost is $2^t+\frac{d}{2}\cdot (192 (\log_2 d-t)-768)$. In this case, the angle precision is $\frac{\alpha}{ 2^{t}}$ where $\alpha$ is an original angle. If $t=0$, then the CNOT-cost is $O(d \log_2d)$ and the original angle is not changed. For $t=\log_2d-4$, the CNOT-cost is $O(d)$ and the angle precision is $\alpha \cdot O(\frac{1}{ d})$.
%
%

One way to reduce the number of CNOT gates dramatically is to pick $\log_2 d +1 $ rotations instead of the full set $d$ and apply them as shown in \cite{kalis18,ziiatdinov2023gaps}.
Those rotations are implemented through $R_y$ gates in \cite{kalis18}, whereas $R_z$ gate is used in many of the real backends \cite{ibmqbackends,ionq} as it is a virtual gate \cite{MWSCG17}. To align with this choice, we adapt the original QFA algorithm to utilize gates from the basic gates set $\{\textup{CNOT}, I, R_z, SX, X \}$, which is also the gate set used by IBM Quantum (IBMQ) backends \cite{ibmqbackends}. We test the new algorithm on a real backend and observe a significant reduction in the depth of the circuit.
%
%

Designing efficient programs tailored for linear nearest neighbor (LNN) architecture, where the multi-qubit gates can be applied only on adjacent qubits has been extensively studied~\cite{saeedi2011synthesis,o2019generalized,bako2022near,zkk2023}. In this paper, we propose an approach to reduce the number of CNOT gates in the circuit for the QFA algorithm when implemented on an LNN architecture. We demonstrate our approach by running the algorithm for $\mathtt{MOD_{37}}$ on the  $ibmq\_manila$ backend. Our implementation results in a circuit with a CNOT-cost that is twice as efficient as the original approach after Qiskit transpilation, and this advantage increases with the length of the input, showing the superiority of our method.  
%
%
%
%
\vspace{-0.5cm}
\section{Preliminaries}
\label{sec:QFAforMODp}
\vspace{-0.3cm}
In this section, we review QFAs and present the QFA algorithm recognizing $\modp$. We refer the reader to \cite{NC00} for the basics of quantum computing and to~\cite{AY21} for further details on QFAs.

\textbf{Quantum finite automaton.} Many different QFA models have been proposed in the literature~\cite{AY21}. The QFA algorithm for $\modp$ is presented by using the most restricted QFA model known, which is also called the Moore-Crutchfield QFA \cite{MC00}; we will refer to this model as QFA throughout the paper. Formally, an $n$-state QFA is a 5-tuple 
$ M=(\Sigma,Q,\{ U_\sigma \mid \sigma \in \Sigma \cup \{\cent,\dollar \}\},q_I,Q_a )$,
 where 
$\Sigma$ is the input alphabet not containing the left and right end-markers (resp., $\cent$ and $\dollar$), $Q = \{q_1,\ldots,q_n\}$ is the set of states, $U_\sigma \in \mathbb{C}^{n \times n}$ is the unitary transition matrix for symbol $\sigma$, $q_I \in Q$ is the initial state, and, $Q_a \subseteq Q$ is the set of accepting state(s).
%
%

From the set of all strings $\Sigma^*$, let $x$ be the given input with $l$ symbols, i.e, $x = x_1 x_2 \cdots x_l$. The computation of $M$ is traced by an $n$-dimensional complex-valued vector called the quantum state. At the beginning of the computation, $M$ is in quantum state $\ket{v_0} = \ket{q_I} $, having zeros except $I$-th entry, which is 1. The input $x$ is processed by $M$ one symbol at a time, by applying sequentially the transition matrix of each scanned symbol including the end-markers. Thus, the final quantum state is 
$
\ket{v_f} = U_\dollar U_{x_l} U _{x_{l-1}} \cdots U_{x_1} U_{\cent} \ket{v_0}.
$ 

After processing the whole input, a measurement in the computational basis is performed. That is, if the $j$-th entry of $\ket{v_f}$ is $a_j$, then the state $q_j$ is observed with probability $|a_j|^2$. Thus, the input $x$ is accepted with the probability of observing an accepting state, i.e., $ \sum_{q_j \in Q_a} |a_j|^2 $.

A language $L \subseteq \Sigma^*$ is recognized by $M$ with bounded error if and only if there exists an error bound $\epsilon \in [0,1/2) $ such that
(i)  for each $x \in L$, $M$ accepts $x$ with probability at least $1 - \epsilon$ and (ii) for each $x \notin L$, $M$ accepts (rejects) $x$ with probability at most $\epsilon$ (at least $1 - \epsilon$).

\textbf{QFA algorithm for $\modp$.} 
In \cite{af98}, it is proven that there exists a QFA with $O(\log p)$ states recognizing the language $\modp$ with bounded error. Before discussing that, we start with a simpler construction. For a fixed  $k \in \{1,\ldots,p-1\}$, let $M_k$ be a QFA with 2 states, namely $\ket{q_1}$ and $\ket{q_2}$. The QFA $M_k$ starts its computation in $\ket{q_1}$ and applies the identity matrix when reading $\cent$ or $\dollar$. When reading symbol $a$, $M_k$ applies a rotation on the (real) plane $\ket{q_1}$-$\ket{q_2}$ with angle $k\frac{2\pi}{p}$. Here, $q_1$ is the accepting state. The QFA $M_k$ can be implemented by using a single qubit by associating $\ket{q_1}$ and $\ket{q_2}$ with states $\ket{0}$ and $\ket{1}$, respectively. Then, the rotation matrix (parameterized with $k$) is
$
R_k = \mymatrix{cc}{\cos \mypar{\frac{2k\pi}{p}} & -\sin \mypar{\frac{2k\pi}{p}} \\ \sin \mypar{\frac{2k\pi}{p}} & \cos \mypar{\frac{2k\pi}{p}}}.$

Note that any member of $\modp$ is accepted with probability 1. On the other hand, the accepting probability for non-members is bounded by $\cos^2 \mypar{\frac{\pi}{p}}$ and thus, $M_k$ can not recognize $\modp$ with bounded error. 

Next, we describe a $2d$-state QFA $M_K$ for the $\modp$ language, where $K=\{k_1,\ldots,k_d\}$. We pick $d$ as a power of 2 for simplicity. The QFA $M_K$ executes the QFAs $M_{k_1}$, \dots, $M_{k_d}$ in parallel. After reading $\cent$, $M_K$ enters into equal superposition of $M_{k_j}$'s for $j \in \{1,\ldots,d\}$. Each $M_{k_j}$ uses two states, and when reading each $a$, $M_{k_j}$ applies the rotation matrix $R_{k_j}$. At the end of the computation, $M_K$ applies the inverse of the transition matrix used for the symbol $\cent$. It was shown in \cite{af98,AN09} that there exist $O(\log p)$ $k$ values such that $M_K$ recognizes $\modp$ with bounded error. (Thus, $d = \Theta(\log p)$.) 

When implementing $M_K$ (on real hardware), we can use Hadamard operators for the end-markers. Consequently, the main cost of the implementation comes from the transition matrix for symbol $a$, which is defined as
$
U_a = \mymatrix{c| c | c | c}{ R_{k_1} & 0 & \cdots & 0 \\ \hline 0 & R_{k_2} & \cdots & 0 \\ \hline \vdots & \vdots & \ddots & \vdots \\ \hline 0 & 0 & \cdots & R_{k_d} } .
$
The operator $U_a$ is also known as uniformly controlled rotation \cite{mottonen2006decompositions}. Thus, we can use its efficient implementations given in the literature, when implementing $M_K$ (Sections \ref{sec:mottonen} and \ref{sec:optimization}).
\vspace{-0.5cm}
\section{Device independent implementations}
\label{sec:independent}
\vspace{-0.4cm}
Here, we compile the known results in the literature and present different generic (i.e., device independent) implementation schema with their costs. We consider three different costs: the number of CNOT gates, the number of single-qubit gates, and the precision of rotation angles. We observe that there may be trade-offs between the CNOT-cost and the precision of rotation angles. In the next section, we discuss how to obtain further improvements if the specification of real hardware is available.

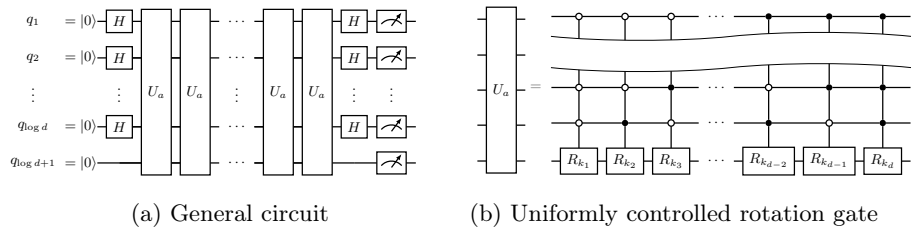
\begin{figure}[h]
\begin{subfigure}{0.5\textwidth}
     \begin{tikzpicture}
\node[scale=0.6] {
   \begin{quantikz}[column sep=0.2cm, row sep = 0.2cm]
        q_{1}&=\ket{0} & \gate{H} & \gate[5, nwires=3]{U_a} & \gate[5, nwires=3]{U_a} &\qw & \cdots &  & \gate[5, nwires=3]{U_a}  & \gate[5, nwires=3]{U_a}& \gate{H} &\meter{} & \qw \\  
        q_{2}&= \ket{0} & \gate{H} & \qw         &\qw  
        &\qw & \cdots & & \qw &\qw &
        \gate{H} &\meter{} & \qw \\
        \vdots & & \vdots  &  \vdots        &  
        &  & \vdots & &  & &\vdots &
        \vdots & \\
    q_{\log d}&= \ket{0} & \gate{H} & \qw         &\qw  
        &\qw & \cdots & & \qw &\qw &
        \gate{H} &\meter{} & \qw \\
    q_{\log d+1}&=\ket{0} & \qw & \qw         &\qw  
        &\qw & \cdots & &       \qw &\qw &
        \qw &\meter{} & \qw 
                \end{quantikz}
                };
\end{tikzpicture}
    \caption{General circuit} \label{fig:qfa}
  \end{subfigure}
\begin{subfigure}{0.45\textwidth}

\begin{tikzpicture}
\node[scale=0.6] {
                \begin{quantikz}[column sep=0.2cm, row sep=0.28cm]
  \ghost{H}\qw & \gate[5]{U_a} & \qw   \\
   \ghost{H}\qw &  & \qw   \\
    \ghost{H}\qw &  & \qw   \\
  \ghost{H} \qw &  & \qw   \\
 \ghost{H}\qw  &  &\qw 
\end{quantikz}=  
    \begin{quantikz}[ column sep=0.2cm, row sep=0.29cm]
  \ghost{H}\qw & \octrl{2} &  \octrl{2}  & \octrl{2}  &\qw & \cdots &&\ctrl{2} &  \ctrl{2}  & \ctrl{2} &\qw\\
  \ghost{H} \wave&&&&&&&&&&\\
  \ghost{H} \qw & \octrl{1} &  \octrl{1}  & \ctrl{1}  &\qw & \cdots &&\octrl{1} &  \ctrl{1}  & \ctrl{1} &\qw\\
  \ghost{H}\qw & \octrl{1} &  \ctrl{1}  & \octrl{1}  &\qw & \cdots &&\ctrl{1} &  \octrl{1}  & \ctrl{1} &\qw\\
 \ghost{H}\qw &\gate{R_{k_1}} & \gate{R_{k_2}} & \gate{R_{k_3}} &\qw & \cdots & & \gate{R_{k_{d-2}}} & \gate{R_{k_{d-1}}} &\gate{R_{k_d}} & \qw
\end{quantikz}
                };
\end{tikzpicture}

  \caption{Uniformly controlled rotation gate} \label{fig:naive}
  \end{subfigure}
  
\caption{QFA construction for $\modp$ language.}
\end{figure}

\begin{figure}[h]
\begin{center}
\begin{tikzpicture}
\node[scale=0.7] {
  \begin{quantikz}[ column sep=0.3cm, row sep=0.2cm]
  \qw & \ctrl{3} & \qw   \\
   \wave&&&\\
 \ghost{H} \qw & \ctrl{1} & \qw   \\
\ghost{H}   \qw & \ctrl{1} & \qw   \\
 \qw  & \gate{R_{\theta}}  &\qw 
\end{quantikz}=  
    \begin{quantikz}[thin lines, column sep=0.3cm, row sep=0.2cm]
  \qw &\qw & \ctrl{3} & \qw & \ctrl{3}  &\qw  \\
   \wave&&&&&\\
  \ghost{H}  \qw &\qw & \ctrl{1} & \qw & \ctrl{1}  &\qw  \\ 
\ghost{H}    \qw &\qw & \ctrl{1} & \qw & \ctrl{1}  &\qw  \\
 \qw &\gate{R_{\theta/2}} & \targ{} & \gate{R_{-\theta/2}} & \targ{} &\qw 
\end{quantikz}
};
\end{tikzpicture}
\end{center}
\caption{Multi-qubit controlled rotation operation decomposition.}
\label{fig:mcrot}
\end{figure}
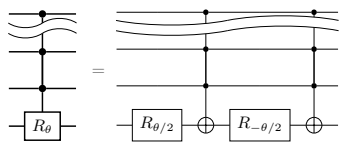

Let $x = a^m$ be the given input. We use $\log d + 1$ qubits to implement $M_K$ (see Fig.~\ref{fig:qfa}). The $\log d $ qubits are used to differentiate $d$ subautomata and the additional qubit is the target qubit. We start in a state $\ket{0}$ for each qubit. We apply Hadamard on the first $\log d$ qubits, and apply identity operator on the last qubit. Thus, we obtain $d$ sub-QFAs by reserving two states for each of them, and any operator applied to a sub-QFA is controlled by the first $\log d$ qubits. For each symbol of $x$, we apply $U_a$. At the end, we again apply Hadamard on the first $\log d$ qubits followed by measurement operators. The input is accepted if and only if all zeros ($\ket{0 \cdots 0}$) is observed. As seen from Fig.~\ref{fig:qfa}, it is enough to find the CNOT-cost of a single $U_a$, i.e., the number of CNOT gates when $U_a$ is decomposed into a set of single-qubit and CNOT gates.
We implement each $R_{k_j}$ of $U_a$ independently as shown in Fig~\ref{fig:naive}. 
The implementation of  $U_a$ uses Gray code for decreasing X gates
\footnote{As the control operators are fired when a qubit is in state $\ket{1}$, we use several NOT ($X$) gates between controlled rotations. For example, we apply NOT gates on the first $\log d$ qubits before and after the controlled $R_{k_1}$ operator, and, in this way, we guarantee that $R_{k_1}$ is fired only if the first $\log d$ qubits are in $ \ket{0 \cdots 0} $. If we follow the order of indices on the circuit, then there will be several NOT gates. But, if we follow an order based on Gray code, then it will be enough to use only a single NOT gate between the controlled rotations.}.
%
We have $d$ multi-qubit controlled rotations in our circuit for $U_a$. Each multi-qubit controlled rotation $R_\theta$ by some angle $\theta$ can be replaced by two multi-qubit controlled NOT gates and two controlled rotation gates  as shown in Fig.~\ref{fig:mcrot} due to \cite{barenco1995elementary}.  The CNOT-cost of multi-qubit controlled NOT gate is bounded by $48(\log d - 3)$ \cite{barenco1995elementary} if one ancilla qubit is available. Thus, the CNOT-cost of $U_a$ is bounded by $48\cdot d(\log d - 3)$.
\vspace{-1cm}
\subsection{Using M\"{o}tt\"{o}nen et al. approach}
\label{sec:mottonen}
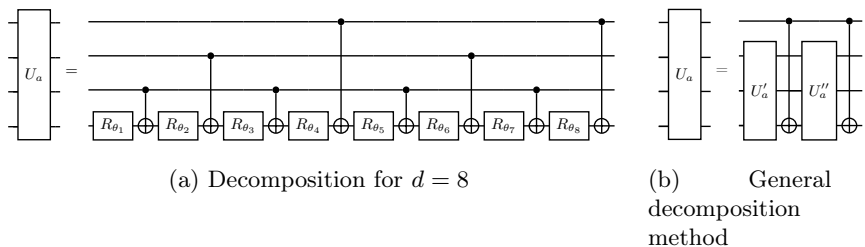
\begin{figure}[h]
\begin{subfigure}[t]{0.70\textwidth}
\begin{tikzpicture}
\node[scale=0.65] {
  \begin{quantikz}[ column sep=0.1cm, row sep = 0.2cm]
  &\ghost{H}\qw & \gate[4]{U_a} & \qw & \qw  \\
  &\ghost{H} \qw & \qw & \qw  & \qw\\
  &\ghost{H} \qw & \qw & \qw  & \qw\\
 &\ghost{H} \qw  & \qw  &\qw & \qw
\end{quantikz} $=$  
    \begin{quantikz}[column sep=0.1cm, row sep = 0.2cm]
   \ghost{R} \qw  &  \qw &  \qw &\qw &  \qw &\qw &  \qw &  \qw &\ctrl{3} &  \qw &  \qw &\qw   &\qw &  \qw &\qw&\qw &\ctrl{3} & \qw &\\
  \ghost{H} \qw &  \qw &  \qw &\qw &  \ctrl{2} &\qw &  \qw &  \qw &  \qw &\qw &  \qw &\qw & \ctrl{2}&  \qw &\qw  &\qw &\qw&\qw \\
 \ghost{H} \qw &  \qw & \ctrl{1} &\qw & \qw &\qw &  \ctrl{1} &  \qw &  \qw &\qw &  \ctrl{1} &\qw &  \qw &\qw & \ctrl{1} & \qw & \qw & \qw &\\
& \gate{R_{\theta_1}} & \targ{} &\gate{R_{\theta_2}} & \targ{} &\gate{R_{\theta_3}} & \targ{} &\gate{R_{\theta_4}} & \targ{} &\gate{R_{\theta_5}} & \targ{}& \gate{R_{\theta_6}} &  \targ{}& \gate{R_{\theta_7}} & \targ{}& \gate{R_{\theta_8}} &  \targ{}& \qw &
\end{quantikz}
};
\end{tikzpicture}
\caption{Decomposition for $d=8$ } \label{fig:uniformm} 
\end{subfigure} 
\begin{subfigure}[t]{0.20\textwidth}
\centering
\begin{tikzpicture}
\node[scale=0.65] {
\begin{quantikz}[column sep=0.2cm, row sep = 0.2cm]
  \ghost{H}\qw & \gate[4]{U_a} & \qw   \\
   \ghost{H}\qw & \qw & \qw   \\
   \ghost{H}\qw & \qw & \qw   \\
 \ghost{H}\qw  & \qw  &\qw
\end{quantikz}=
\begin{quantikz}[column sep=0.1cm, row sep = 0.17cm]
  \ghost{H}\qw &\qw & \ctrl{3} & \qw  & \ctrl{3} & \qw    \\
 \ghost{H} \qw & \gate[3]{U_a'}  & \qw  & \gate[3]{U_a''} & \qw  & \qw   \\
  \ghost{H}\qw &  \qw & \qw  &  \qw  & \qw  & \qw   \\
 \ghost{H}\qw  &  \qw   &\targ{} &  \qw   &\targ{}&\qw 
\end{quantikz}
};
\end{tikzpicture}
\caption{General decomposition method }\label{fig:recmot}
\end{subfigure}
\caption{Efficient decomposition  of the uniformly controlled rotation gate \cite{mottonen2006decompositions}.}
\end{figure}

M\"{o}tt\"{o}nen et al. presented an efficient decomposition of the uniformly controlled rotation gate (see Fig.~\ref{fig:uniformm}) in  \cite{mottonen2006decompositions}. This circuit is based on recursive application of the circuit given in Fig.~\ref{fig:recmot}. They proposed a construction for $U_a$ consisting of $d$ CNOT gates and $d$ rotation gates.

Let $\alpha_i=2 \pi k_i/p$ be the original angle that is used in $M_{k_i}$ for $i \in \{1, \ldots, d\}$. 
The angles $\{\theta_i\}$ from the efficient decomposition of M\"{o}tt\"{o}nen et al. can be obtained from  $\{\alpha_i\}$ by using the equation
$\begin{pmatrix}
       \theta_1 & \theta_2 
    & \cdots & \theta_d 
\end{pmatrix}^T
= B 
\begin{pmatrix} \alpha_1 & \alpha_2 & \cdots & \alpha_{d}
\end{pmatrix}^T $,
where $ B_{ij}=\frac{1}{d}(-1)^{(b_{j-1}) \cdot (g_{i-1})}$ and $b_m$ and $g_m$ are binary code and binary reflected Gray code representations of the integer $m$, respectively.
Thus, the CNOT-cost of $U_a$ is bounded by $d$, but the angles for the target qubit rotations are transformed, using more sensitive rotation angles, i.e., we use $O(\frac{2\pi}{p\log p})$ instead of $O(\frac{2\pi}{p})$.
\vspace{-0.2cm}
\subsection{Optimization}
\label{sec:optimization}

We already observed a better CNOT-cost in Section~\ref{sec:mottonen} but at the cost of using more sensitive rotation of angles. Remark that, for the current quantum computers, there is a limit to the accuracy of rotation implementing up to angles $\theta+0.0001$ \cite{maldonado2022error}. Our main goal is to optimize a given quantum circuit for uniformly controlled rotation operations with respect to not only the CNOT-cost but also the precision of rotation angles.

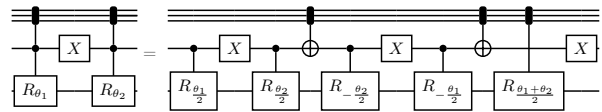
\begin{figure}[h]
\begin{center}
\begin{tikzpicture}
\node[scale=0.7] {
\begin{quantikz}[column sep=0.05cm, row sep = 0.2cm]
\qw &\ctrlbundle{1} & \qwbundle[alternate]{}  &\ctrlbundle{1}  & \qw \qwbundle[alternate]{} \\
\qw & \ctrl{1} & \gate{X}  & \ctrl{1} & \qw \\
\ghost{R_{\frac{\theta_1+\theta_2}{2}}} \qw  & \gate{R_{\theta_1}}  & \qw  & \gate{R_{\theta_2}}  & \qw 
\end{quantikz}=  
\begin{quantikz}[column sep=0.05cm, row sep = 0.2cm]
\qwbundle[alternate]{}  &  \qwbundle[alternate]{}  &  \qwbundle[alternate]{}  &  \qwbundle[alternate]{}  & \ctrlbundle{1}  &   \qwbundle[alternate]{}  &  \qwbundle[alternate]{}  &  \qwbundle[alternate]{}  & \ctrlbundle{1}  &   \ctrlbundle{2}  &  \qwbundle[alternate]{}  &  \qwbundle[alternate]{} \\
    \qw & \ctrl{1} &\gate{X} & \ctrl{1}& \targ{} & \ctrl{1} &\gate{X} & \ctrl{1}& \targ{} &\qw &\gate{X}  & \qw  \\
 \qw &\gate{R_{\frac{\theta_1}{2}}} &  \qw &\gate{R_{\frac{\theta_2}{2}}} & \qw &\gate{R_{-\frac{\theta_2}{2}}} &  \qw &\gate{R_{-\frac{\theta_1}{2}}} &\qw &\gate{R_{\frac{\theta_1+\theta_2}{2}}} & \qw  & \qw 
\end{quantikz}
};

\end{tikzpicture}
\end{center}
\caption{Proposed decomposition for a pair of rotations.}
\label{fig:app-pairss}
\end{figure}

\begin{figure}[h]
\begin{subfigure}{0.5\textwidth}
\centering
\begin{tikzpicture}
\node[scale=0.7] {
\begin{quantikz}[column sep=0.2cm, row sep = 0.2cm]
\ghost{H}\qwbundle[alternate]{}  &\ctrlbundle{1} & \qwbundle[alternate]{}  \\
\ghost{H}\qw & \ctrl{1} & \qw \\
\ghost{R_{\frac{\theta}{2}}} \qw  & \gate{R_{\theta}}  & \qw 
\end{quantikz}=  
\begin{quantikz}[column sep=0.2cm, row sep = 0.2cm]
\ghost{H}  \qwbundle[alternate]{}  &  \qwbundle[alternate]{}  & \ctrlbundle{1}  &   \ghost{H}\qwbundle[alternate]{}  &  \ctrlbundle{1}  &   \ctrlbundle{2}  &  \qwbundle[alternate]{} \\
  \ghost{H}   \qw & \ctrl{1} & \targ{} & \ctrl{1}& \targ{} &\qw & \qw\\
 \ghost{H} \qw &\gate{R_{\frac{\theta}{2}}} & \qw &\gate{R_{-\frac{\theta}{2}}}  &\qw &\gate{R_{\frac{\theta}{2}}} & \qw
\end{quantikz}
};
\end{tikzpicture}
\caption{Multi-qubit controlled rotation}
\label{fig:multicontrR}
\end{subfigure}
\begin{subfigure}{0.5\textwidth}
\centering
\begin{tikzpicture}
\node[scale=0.5] {
\begin{quantikz}[column sep=0.2cm, row sep = 0.2cm]
 \ghost{H}\qwbundle[alternate]{}  &\ctrlbundle{1} & \qwbundle[alternate]{} \\
 \ghost{H}\qw & \ctrl{1} & \qw \\
 \ghost{H}\qw & \ctrl{2} & \qw \\
\ghost{H} \qw & \qw & \qw \\
 \ghost{H}\qw  & \targ{}  & \qw 
\end{quantikz}=  
\begin{quantikz}[column sep=0.1cm, row sep = 0.2cm]
  \ghost{H} \qwbundle[alternate]{}  &  \ctrlbundle{3}  &   \qwbundle[alternate]{}  &  \ctrlbundle{3}  & \qwbundle[alternate]{} & \qwbundle[alternate]{}   \\
    \ghost{H} \qw &  \qw &\ctrl{1} &  \qw &\ctrl{1} &  \qw \\
  \ghost{H}\qw &  \qw &\ctrl{1} &  \qw &\ctrl{1} &  \qw \\
 \ghost{H}\qw & \targ{} & \ctrl{1}  & \targ{} & \ctrl{1} \qw & \qw \\
 \ghost{H} \qw & \qw & \targ{} & \qw & \targ{} & \qw & \qw 
\end{quantikz}
};
\end{tikzpicture}
\caption{Multi-qubit controlled NOT}
\label{fig:multicontrNOT}
\end{subfigure}
\caption{Multi-qubit controlled gate decompositions from \cite{barenco1995elementary}.}
\end{figure}
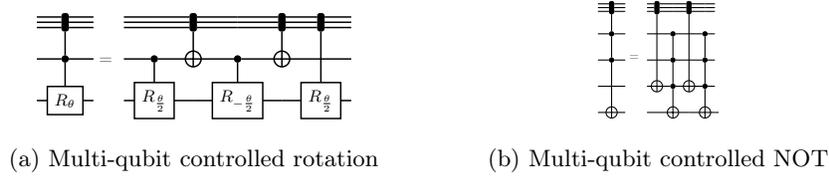

First, we propose a decomposition of a circuit implementing a pair of multi-qubit controlled rotations as given in Fig.~\ref{fig:app-pairss}. This decomposition is based on a known decomposition of a single multi-qubit controlled rotation operation proposed in \cite{barenco1995elementary} and shown in Fig.~\ref{fig:multicontrR}. 
In our case, we implement consequently two multi-qubit controlled rotations $R_{\theta_1}$ and $R_{\theta_2}$, where one of the controller qubits is in orthogonal states for each rotation. That means we apply NOT gate between controlled rotations to this control qubit. For the operation with $n$ control qubits, without loss of generality, we assume that this control qubit is the $n^{th}$ qubit in the register and the target qubit is the $(n+1)^{th}$ qubit. 

The CNOT-cost of this implementation is at most $4 \cdot 48 (n-4) = 192(n-4)$. The cost is based on the circuit for $g$-qubit controlled NOT gate presented in \cite{barenco1995elementary}, where $g>3$. This operator can be implemented, as long as a single qubit is available to use as ancilla. The CNOT-cost of this  $g$-qubit controlled NOT gate is bounded by $ 48(g-3)$. 

On the right circuit in Fig.~\ref{fig:app-pairss}, each one of the two multi-qubit controlled NOT gates is $(n-1)$-qubit controlled. The target $(n+1)^{th}$ qubit is an ancilla qubit for the 2 multi-qubit controlled NOT gates. An $(n-1)$-controlled rotation gate is  decomposed into 2 rotation gates and 2 multi-qubit controlled NOT gates as shown in Fig.~\ref{fig:mcrot}. For their decomposition, we use the $n^{th}$ qubit as an ancilla.   
Thus, the CNOT-cost of the decomposed circuit (on the right hand side) in Fig.~\ref{fig:app-pairss}
is bounded by $4 \cdot 48 (n-4) = 192(n-4)$.
The naive implementation of the circuit on the left hand side of  Fig.~\ref{fig:app-pairss} requires 2 decomposition of $n$-controlled rotation gates. The CNOT-cost of such implementation is bounded by $4 \cdot 48 (n-3) = 192(n-3)$.

Then, our optimization approach is to combine the methods presented in Fig.~\ref{fig:recmot} and Fig.~\ref{fig:app-pairss}. 
To achieve better precision of rotations, we can reduce the number of iterations in the method given in Fig.~\ref{fig:recmot}. For instance, after $t$ iterations of the decomposition, the angles are reduced up to $2^{t}$ times. We can stop recursion at this step. Then, the rest $2^t$ undecomposed uniformly controlled rotations can be split based on the pairs of rotations as in Fig.~\ref{fig:app-pairss}.
Compared to the original circuit, the angles are reduced by $2^{t+1}$ times. For $ t<\log_2 d-4$, the CNOT-cost is $2^t+\frac{d}{2}\cdot (192 (\log_2 d-t)-768)$.
Namely, after $t$ iterations, the circuit consists of $2^t$ CNOT gates and $2^t$  undecomposed  uniformly controlled rotation gates, where each of them consists of $2^{\log_2 d-t-1}$ pairs of $\log_2 d-t$ - controlled rotations.

The same precision can be achieved if we implement $t+1$ recursive iterations and use naive circuit (Fig.~\ref{fig:naive}) for the rest $2^{t+1}$ undecomposed  uniformly controlled rotation gates.  In this case, the CNOT-cost is 
$2^{t+1}+ \frac{d}{2}\cdot (192(\log_2 d-t-1)-576)=2\cdot 2^t+\frac{d}{2}(192(\log_2 d-t)-768)$,
which means that the advantage of the optimized circuit over the naive circuit is $2^t$ CNOT gates. 

Finally, we obtain a bound for CNOT-cost between $d$ and $d \log d$ with  rotation angles better than $ O(\frac{2\pi}{p\log p}) $.
\vspace{-0.4cm}
\subsection{Pseudo rotations approach}
\label{sec:pseudo}
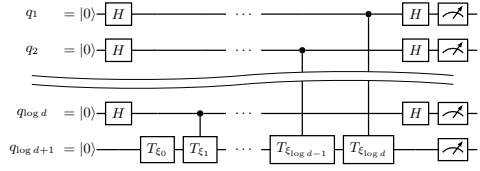
\begin{figure}[h]
\begin{center}
\begin{tikzpicture}
\node[scale=0.6] {
\begin{quantikz}[column sep=0.2cm, row sep=0.2cm]
            q_{1}&=\ket{0} & \gate{H} & \qw & \qw  &\qw & \ldots && \qw   & \ctrl{6}& \gate{H} &\meter{} & \qw \\
           q_{2}&=\ket{0} & \gate{H}& \qw  & \qw &\qw &\ldots && \ctrl{5}&\qw     &  \gate{H} &\meter{} & \qw \\
           \\
           & & \wave &&&&&&&&& \\
            & & \\
                q_{\log d}&= \ket{0} & \gate{H} & \qw & \ctrl{1}   &\qw & \ldots && \qw  &  \qw &\gate{H} &\meter{} & \qw \\
                q_{\log d+1}&=\ket{0} & \qw & \gate{T_{\xi_0}} &  \gate{T_{\xi_1}}  & \qw & \ldots  && \gate{T_{\xi_{\log d-1}}} &   \gate{T_{\xi_{\log d}}} &\qw & \meter{}& \qw
\end{quantikz}
};
\end{tikzpicture}
\end{center}
\caption{Pseudo rotations approach.}
\label{fig:pseudo}
\end{figure}

\vspace{-0.2cm}
The operator $U_a$ is composed of $d$ rotations. One way to reduce the number of CNOT gates dramatically is to pick $\log_2 d +1 $ rotations and apply them as shown in Fig.~\ref{fig:pseudo} \cite{kalis18}. Even though the CNOT-cost is $O(\log d)$ in this case, each $R_{k_j}$ will be a linear combination of some rotations, and so the set $K$ is not determined independently, i.e., most choices of $K$ cannot be constructed with this method. Namely, if the state of control qubits is $|c\rangle=|c_1c_2 \ldots c_{\log d}\rangle$ then the target qubit is rotated by an angle $\xi=\xi_0+c_1\xi_1+c_2 \xi_2 + \cdots + c_{\log d}\xi_{\log d}$. This $\xi$ is should be equal to $\frac{2k_c\pi}{p}$, where $c_1c_2\ldots c_{\log d}$ is a binary representation of $c$. It means that the set $K$ of size $d$ is constructed by a linear combination of $\log d+1$ angles. It does not work for all cases. However, numerical experiments given in \cite{ziiatdinov2023gaps}, show that there exists such $\xi$ that produces the required set $K$ which allows to recognize $\modp$ with bounded error at most $\frac{1}{3}$. 

\vspace{-0.4cm}
\section{Implementations on real hardware}
\label{sec:real-hardware}
\vspace{-0.3cm}
Any program sent for execution on a real quantum computer is transpiled into some sequences of basis gates. Each computer may have a different set of basis gates. Thus, re-writing or modifying the program based on the basis gates can reduce the number of gates in the program, as the automatic transpilation processes are usually generic and so may not guarantee the best solutions. 
Similar to knowing the basis gates of a real hardware, its qubit topology may also help us to reduce especially the CNOT-costs. For example, a CNOT gate between two qubits that are not neighbour to each other would be implemented by using several CNOT gates.

In this section, we show how we can reduce the number of gates further based on the choice of basis gates and the topology of the real hardware.  For the implementation, we use tools and libraries provided by Qiskit framework~\cite{qiskit_framework}. For each case, we pick one real hardware of IBM Quantum (IBMQ) \cite{ibmqbackends}. 
Almost all IBMQ backends use the set of gates $\{CX, I, R_z, SX, X \}$ as basis gates. Here, $CX$ is the CNOT gate, $I$ is the identity gate, $X$ is the NOT gate, and $R_z$ and $SX$ are defined as:
$	R_z(\theta) = \mymatrix{cc}{e^{-i\theta/2} & 0 \\ 0 & e^{i \theta/2}},
~~
	SX =  \frac{1}{2} 
	\begin{pmatrix}
        1+i & ~ & 1-i \\
        1-i &~ & 1+i \\
    \end{pmatrix}.$

\textbf{Re-writing programs based on basis gates } 
A single quantum bit is modelled by the Bloch sphere \cite{NC00}. The rotation operator $R_{k_j}$ is defined between $\ket{0}$ and $\ket{1}$ as a real-valued operator, denoted usually $R_y$:
$
	R_y(\theta) = \mymatrix{cc}{\cos \theta & -\sin \theta \\ \sin \theta & \cos \theta}.$ 
On the other hand, almost all IBMQ backends use $R_z$ gate as the basic rotation operator as mentioned before. Hence any $R_y$ gate should be expressed in terms of $R_z$ and other gates in the basis gate set. 

To comply with the basis gate set of IBMQ, we will use $R_z$ gates instead of $R_y$ gates in the QFA algorithms for $\modp$. For simplicity, we pick $\theta = 2\pi/p$. We start with the following lemma. 

\noindent
\textbf{Lemma 1.}		$SX^{\dagger} R_z(2\pi/p) SX = R_y(2\pi/p)$. 

\begin{proof}
\label{app:sxua}

We can re-write $SX$ as follows:
\begin{equation}
 \frac{1}{2} \begin{pmatrix}
        1+i & ~ & 1-i \\
        1-i &~ & 1+i \\
    \end{pmatrix}
    =
    \frac{1}{\sqrt{2}}
    \begin{pmatrix}
        e^{i \pi / 4} & ~ & e^{-i\pi / 4} \\
        e^{-i\pi /4} &~ & e^{i \pi /4} \\
    \end{pmatrix}
	=
    \frac{1}{\sqrt{2}}
    e^{-i \pi /4}
    \begin{pmatrix}
        e^{i \pi /2} & ~ & 1 \\
        1 &~ & e^{i \pi /2} \\
    \end{pmatrix}.
\end{equation}
\noindent 
We calculate $ SX^{\dagger} R_z(2\pi/p) SX$:
\begin{align}
    & =
    \frac{1}{\sqrt{2}} e^{i \pi / 4} \mymatrix{cc}{ e^{-i \pi /2} & 1 \\ 
    1 &  e^{-i \pi /2}
    }
    \mymatrix{cc}{ e^{-2\pi i / p } & 0 \\ 0 & e^{ 2\pi i /p} }
    \frac{1}{\sqrt{2}} e^{-i \pi / 4}
    \begin{pmatrix}
        e^{i\pi / 2} & ~ & 1 \\
        1 &~ & e^{i \pi / 2} \\
    \end{pmatrix}  \nonumber
    \\    & 
    = \frac{1}{2}e^0 
    \mymatrix{cc}{ 
    e^{-i \pi /2} & 1 \\
    1 &  e^{-i \pi /2}
    }
    \mymatrix{cc}{
    e^{ i \mypar{ {\pi / 2}- {2\pi / p} } } & 
    e^{ -i 2\pi / p } \\
    e^{ i 2\pi / p } &
    e^{ i \mypar{ {\pi / 2}+{2 \pi /p} } }
    } \nonumber 
    \\ & 
    = \frac{1}{2}
    \mymatrix{ccc}{
        e^{ -i 2\pi / p } + e^{ i 2\pi /p }  & ~~~~ &
        e^{ -i \mypar{ {\pi /2} + {2\pi / p}  } } +
        e^{ i \mypar{ {\pi / 2} + {2\pi /p}  } } \\
        e^{ i \mypar{ {\pi /2} - {2\pi /p}  } } +
        e^{ -i \mypar{ {\pi /2} - {2\pi /p}  } } & ~~~~ &
        e^{ -i 2\pi / p } + e^{ i 2\pi / p }
    }.
\end{align}

We calculate each term of this matrix one by one by using the following trigonometric qualities: (i) $ \cos(-\theta) = \cos(\theta) $, (ii) $ \sin(-\theta) = -\sin(\theta) $, (iii) $ e^{i\theta} + e^{-i\theta} = 2\cos(\theta) $, and (iv) $ \cos (\theta+\pi/2) = -sin(\theta) $.

The top-left and the bottom-right terms are equal to
\begin{equation}
    e^{ -i 2\pi / p } + e^{ i 2\pi / p } = 2 \cos \mypar{ 2\pi / p }.
\end{equation}
The top-right term is equal to
\begin{align}
     e^{ -i \mypar{ {\pi / 2} + {2\pi/p}  } } +
    e^{ i \mypar{ {\pi / 2} + {2\pi / p}  } } 
    &= 
    2\cos \mypar{ {\pi / 2} + {2\pi / p} }  \nonumber \\ 
    &= -2 \sin \mypar{ 2\pi / p } .
\end{align}
The bottom-left term is equal to 
\begin{align}
    e^{ i \mypar{ {\pi/2} - {2\pi/p}  } } +
        e^{ -i \mypar{ {\pi / 2} - {2\pi / p}  } } &=  2\cos \mypar{ {\pi/2} - {2\pi/p}} \nonumber \\ 
        &= -2 \sin \mypar{ - 2\pi / p } \nonumber \\ 
        &= 2 \sin \mypar{ 2\pi / p }.
\end{align}
Thus, we obtain that $SX^{\dagger} R_z(2\pi/p) SX$ is equal to
\[
    \mymatrix{ccc}{
     cos(2\pi / p) & ~~ & -sin(2\pi / p) \\
     sin(2\pi / p) & ~~ & cos(2\pi / p)
     }
     = R_y(2\pi/p).
\]
\qed
\end{proof}

We can easily obtain the followings.

\noindent
\textbf{Corollary 1.}	$SX^{\dagger} R_z(2k\pi/p) SX = R_y(2k\pi/p)$.

\noindent
\textbf{Corollary 2.}	$ (R_y(2k\pi/p))^j = (SX^{\dagger} R_z(2k\pi/p) SX )^j = SX^{\dagger} R^j_z(2k\pi/p) SX  $.

Therefore, for the 2-state QFA, we can apply $ SX $ and $SX^{\dagger}$ on the left and right end-markers instead of Hadamard gate. For each symbol, we apply $ R_z(2k\pi/p) $ for some $0<k<p$.
For the $2d$-state QFA (defined in Section~\ref{sec:QFAforMODp}), we make the following modifications. For each $ R_{k_j} $ defined with $R_y$, we define $ R'_{k_j} $ defined with $ R_z $, where the rotation angles are identical. When reading $a^j$, the computation of original algorithm is represented as
\[
	\underbrace{\mypar{H^{\otimes \log d} \otimes I} }_{U_\dollar}
	\underbrace{
	\mymatrix{c| c | c | c}{ R_{k_1} & 0 & \cdots & 0 \\ \hline 0 & R_{k_2} & \cdots & 0 \\ \hline \vdots & \vdots & \ddots & \vdots \\ \hline 0 & 0 & \cdots & R_{k_d} }^j}_{U^j_a}
	\underbrace{ \mypar{ H^{\otimes \log d} \otimes I} }_{U_{\cent}}.
\] 
The computation of the modified algorithm is as follows:
\[
	\underbrace{
	U_\dollar \mymatrix{c c c c}{
        SX^{\dagger} & 0 & \cdots & 0 \\
        0 & SX^{\dagger}& \cdots & 0 \\
        \vdots & \vdots &  \ddots & \vdots \\
        0 & 0 & \cdots & SX^{\dagger}
    }
    }_{U'_\dollar}
    \underbrace{
    \mymatrix{c c c c}{
        R'_{k_1} & 0 & \cdots & 0 \\
        0 & R'_{k_2} & \cdots & 0 \\
        \vdots & \vdots &  \ddots & \vdots \\
        0 & 0 & \cdots & R'_{k_d}
    }^j 
    }_{(U'_a)^j}
    \underbrace{
    \mymatrix{c c c c}{
        SX & 0 & \cdots & 0 \\
        0 & SX & \cdots & 0 \\
        \vdots & \vdots &  \ddots & \vdots \\
        0 & 0 & \cdots & SX
    } U_{\cent}
    }_{U'_{\cent}}.
\]

Taking the pseudo rotations approach discussed in Section~\ref{sec:pseudo}, we consider the QFA for the language $\mathtt{MOD_{11}}$ whose transition matrices are defined as above. The circuit consists of three qubits. Note that a classical circuit with the same number of qubits can not recognize $\mathtt{MOD_{p}}$ for $p>8$.  For the input $a^{11}$, the modified algorithm for $\mathtt{MOD_{11}}$ produces a shorter circuit. Namely, the modified circuit has 6 $SX$, 44 $CX$, 44 $R_Z$ gates, the depth of the circuit is 84,  while the original circuit has 70 $SX$, 44 $CX$, 74 $R_Z$ gates and the depth of the circuit is 183. Notably, the modification does not reduce the number of CNOT gates, but we can have shorter circuits by using less $SX$ and $R_z$ gates. Here, we run the algorithms in $\mathtt{ibmq}\_\mathtt{belem}$ backend  \cite{ibmqbackends} and use the default transpilation by Qiskit for both cases.
%
%
%
We refer the reader to the technical report \cite{SalehiY2021} for the further details about this subsection.

\begin{figure}[h!]
\centering
\includegraphics[width=5cm]{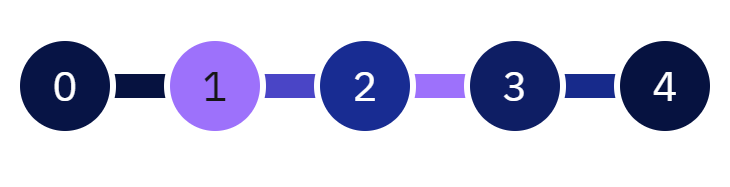}
\caption{Qubits' topology of  $\mathtt{ibmq}\_\mathtt{manila}$ machine.}
\label{fig:manila}
\end{figure}

\textbf{Re-writing programs based on the qubit topology }
IBM Quantum has several backends with linear topology of qubits. For instance, $\mathtt{ibmq}\_\mathtt{manila}$ machine works with 5 linearly connected qubits indexed by 0-1-2-3-4 (see Fig\ref{fig:manila}). When a quantum program is executed on  $\mathtt{ibmq}\_\mathtt{manila}$, the program is transpiled based on not only the basis gates but also its topology. Let us note that the transpiler provided by Qiskit framework optimizes the circuits to produce less computational error \cite{transpiler,corcoles2021exploiting}. 

We implemented the QFA algorithm with pseudo rotations given in Section~\ref{sec:pseudo} for $\mathtt{MOD_{37}}$ on the simulator of  $\mathtt{ibmq}\_\mathtt{manila}$ machine using $\mathtt{FakeManilaV2}$ backend. As in the previous section, we set $p=37$ so that $p>2^5$ because a classic computer using 5 bits can recognize $\modp$ for any $p \leq 32$ \cite{af98}.  Based on the topology, we can try to minimize the number of CNOT gates. For this purpose, we move the target qubit to make it closer to the control qubit for CNOT gate implementation. This procedure differs from the one where we move control qubits closer to the target one. 
Trivially, such moves are made by swapping the target qubit with the neighboring qubits. This approach gives us better results compared to the results reached by the Qiskit transpiler. For comparison, we transpiled the pseudo circuit with optimization level 3. 

\begin{figure}[h!]

\centering
\includegraphics[width=7cm]{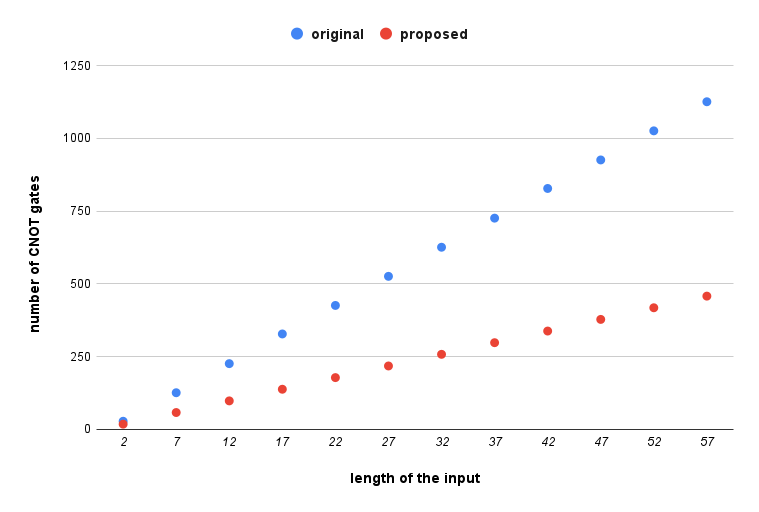}
\caption{The number of CNOT gates after transpilation for the pseudo circuit constructed using the original approach and our approach for $\mathtt{MOD_{37}}$.}
\label{fig:cxs}
\end{figure}

\begin{figure}[h!]
\begin{center}
\begin{tikzpicture}
\node[scale=0.55] {
\begin{quantikz}[column sep=0.2cm, row sep=0.2cm]
            q_{1}=\ket{0} & \gate{H} & \qw  &\qw & \qw  & \qw& \qw & \qw & \qw  & \qw& \qw & \qw & \qw  & \qw& \qw & \qw & \qw   & \ctrl{1}& \qw   & \ctrl{1}&\gate{H} &\meter{} & \qw \\            
           q_{2}=\ket{0} & \gate{H}& \qw  &\qw  & \qw &\qw  & \qw &\qw  & \qw &\qw  & \qw &\qw  & \ctrl{1}&\qw & \targ{}  & \ctrl{1}   & \gate{R_z(\theta_4)} & \targ{} &\gate{R_z(\theta_4')} & \targ{} & \gate{S^{\dagger}}&\meter{} & \qw \\
           q_{3}=\ket{0} & \gate{H}&\qw  & \qw  & \qw &\qw  & \qw  & \ctrl{1}&\qw & \targ{}  & \ctrl{1}   & \gate{R_z(\theta_3)} & \targ{} &\gate{R_z(\theta_3')} & \ctrl{-1} &\targ{} & \qw    & \qw    & \qw& \qw &\gate{H} &\meter{} & \qw \\
           q_{4}=\ket{0} &\gate{S}&\gate{R_z(\theta_1)}&  \targ{} &\gate{R_z(\theta_1')} & \targ{} &\gate{R_z(\theta_2)} &\targ{} &\gate{R_z(\theta_2')} &\ctrl{-1}   & \targ{} &  \qw  & \qw &\qw  & \qw & \qw  & \qw &\qw  & \qw & \qw &\gate{H} &\meter{} & \qw \\
           q_{5}=\ket{0} &\gate{H} &\qw  &\ctrl{-1} &\qw &\ctrl{-1}  &  \qw  & \qw &\qw  & \qw & \qw  & \qw &\qw  & \qw &\qw  & \qw    & \qw& \qw & \qw & \qw &\gate{H} &\meter{} & \qw \\    
\end{quantikz}
};
\end{tikzpicture}
\end{center}
\caption{Proposed circuit for the input $a$ for $\mathtt{MOD_{37}}$ language}.
\label{fig:pseudodec}
\end{figure}
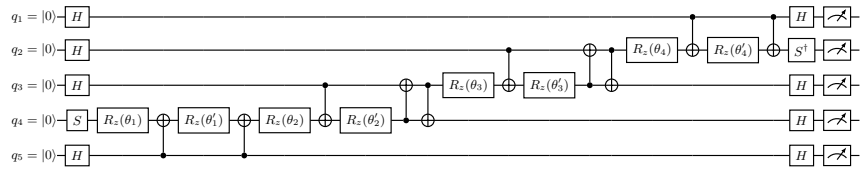

We implemented both our circuit and the pseudo circuit by picking $K=[3, 6, 19, 2, 8]$ due to \cite{ziiatdinov2023gaps}. The results are given in Figure~\ref{fig:cxs}, and we can observe that we reduce the number of CNOT gates by more than half. We present our circuit in Figure~\ref{fig:pseudodec}. Here, we use 5 qubits and set the $4^{th}$ qubit $q4$ as the target, initially. We may also choose qubit $q2$ as the target qubit initially because the positions of $q2$ and $q4$ correspond to an optimized number of swaps of the target qubit with adjacent qubits while moving the target qubit to control qubit. We note that the number of CNOT gates does not change after using Qiskit transpiler with optimization level 3.

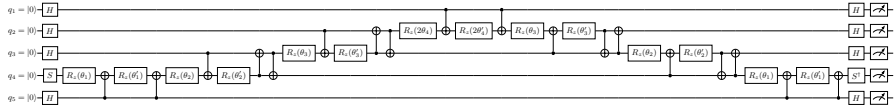
\begin{figure}[h!]
\begin{center}
\begin{tikzpicture}
\node[scale=0.35] {
\begin{quantikz}[column sep=0.2cm, row sep=0.2cm]
            q_{1}=\ket{0} & \gate{H} &\qw & \qw & \qw  & \qw& \qw & \qw & \qw  & \qw& \qw & \qw & \qw  & \qw& \qw & \qw & \qw   & \ctrl{1}& \qw   & \ctrl{1}&  \qw & \qw  & \qw& \qw & \qw & \qw  & \qw& \qw & \qw & \qw  & \qw& \qw& \qw  & \qw &\gate{H} &\meter{} & \qw \\
            
           q_{2}=\ket{0} & \gate{H}& \qw &\qw  & \qw &\qw  & \qw &\qw  & \qw &\qw  & \qw &\qw  & \ctrl{1}&\qw & \targ{}  & \ctrl{1}   & \gate{R_z(2\theta_4)} & \targ{} &\gate{R_z(2\theta_4')} & \targ{}
           &\gate{R_z(\theta_3)} 
           &\targ{} 
           &\gate{R_z(\theta_3')}
           &\ctrl{1}
           &\targ{} 
           &\qw&\qw & \qw& \qw & \qw &\qw&\qw& \qw  & \qw&\gate{H}&\meter{} & \qw \\
           q_{3}=\ket{0} & \gate{H}&\qw & \qw  & \qw &\qw  & \qw  & \ctrl{1}&\qw & \targ{}  & \ctrl{1}   & \gate{R_z(\theta_3)} & \targ{} &\gate{R_z(\theta_3')} & \ctrl{-1} &\targ{} & \qw    & \qw    & \qw& \qw& \qw 
           & \ctrl{-1}  & \qw &\targ{}
           & \ctrl{-1}  & \gate{R_z(\theta_2)} &\targ{}
           &\gate{R_z(\theta_2')} 
           & \ctrl{1}   &\targ{}
            &\qw  & \qw& \qw  & \qw
           &\gate{H} &\meter{} & \qw \\
           q_{4}=\ket{0}  &\gate{S}&\gate{R_z(\theta_1)}&  \targ{} &\gate{R_z(\theta_1')} & \targ{} &\gate{R_z(\theta_2)} &\targ{} &\gate{R_z(\theta_2')} &\ctrl{-1}   & \targ{} &  \qw  & \qw &\qw  & \qw & \qw  & \qw &\qw  & \qw & 
           \qw  & \qw &\qw  & \qw & \qw  & \qw &\qw   &
           \ctrl{-1}   & 
           \qw &\targ{} &\ctrl{-1}   &\gate{R_z(\theta_1)} & \targ{} &\gate{R_z(\theta_1')} &\targ{} &\gate{S^{\dagger}}&\meter{} & \qw \\
           q_{5}=\ket{0} &\gate{H} &\qw &\ctrl{-1} &\qw &\ctrl{-1}  &  \qw  & \qw &\qw  & \qw & \qw  & \qw &\qw  & \qw &\qw  & \qw    & \qw& \qw & \qw & \qw &\qw  & \qw &\qw  & \qw & \qw  & \qw &\qw  & \qw &\qw  & \qw & \qw &\ctrl{-1} &\qw &\ctrl{-1}  &\gate{H} &\meter{} & \qw \\
\end{quantikz}
};
\end{tikzpicture}
\end{center}
\caption{Proposed circuit for the input $aa$ for $\mathtt{MOD_{37}}$ language.} 
\label{fig:pseudodec2}
\end{figure}

Although QFA reads one symbol at a time by definition, When reading more than one input symbol is allowed, we can further get a small improvement. For the input $aa$, we give our circuit in Figure~\ref{fig:pseudodec2}. For each input symbol, the circuit is split into sub-circuits. For the $i$-th symbol, the last controlled rotation of the corresponding sub-circuit is joined with a rotation of the sub-circuit corresponding to the $(i+1)$-th symbol. This simple maneuver reduces one CNOT gate. We apply this idea for all sub-circuits. But, for the last sub-circuit, we should apply one more controlled rotation with the inverse angle for the correction, and this is implemented while reading the right end-marker.

We calculate the number of CNOT gates of our program executed on a device having linearly connected $n$ qubits. For each input symbol, the first controlled rotation requires 2 CNOT gates. Then, 3 CNOT gates are used for each rotation. The total number of rotations is $n-3$. While reading the right end-marker, the last rotation uses 2 CNOT gates.
\vspace{-0.2cm}
\begin{theorem}
	For the input string $ a^j $, the circuit proposed in this subsection for recognizing $\modp$ language uses $(2+3(n-3))j+2  $ CNOT gates when executed on an $n$-qubit device with linear-nearest neighbor topology.    
\end{theorem}

\vspace{-0.6cm}

\section{Experimental results}
\label{sec:experiment}
\begin{figure}[h!]

\centering
\includegraphics[width=11cm]{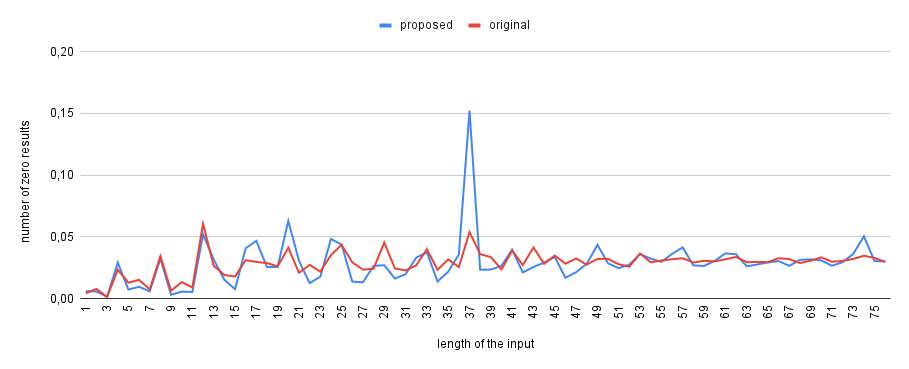}
\caption{The experimental results of the original and the proposed pseudo circuits implementations for recognition $\mathtt{MOD_{37}}$ language.}
\label{fig:exper}
\end{figure}

We also experimentally observed that our approach can give better results. We implemented the original and our proposed circuits on a simulator of $\mathtt{ibmq}\_\mathtt{cairo}$  machine using $\mathtt{FakeCairoV2}$ backend. Unfortunately, 5-qubits  $\mathtt{ibmq}\_\mathtt{manila}$ was not available.  We
applied transpilation with optimization level 3 to both circuits. $\mathtt{ibmq}\_\mathtt{cairo}$ machine has 27 qubits, but we use only 5 qubits in LNN architecture. The set of basic gates is the same as $\mathtt{ibmq}\_\mathtt{manila}$  machine uses. The transpiler maps five virtual qubits to five linearly connected physical qubits while implementing a proposed circuit. The results are shown in Fig.\ref{fig:exper}, where the y-axis shows rates of
acceptances of input words. For each input with length $ 1,2,3,\ldots,75 $, both circuits were executed 100 000 times.

The red and blue lines show the results for the original and optimized circuits, respectively. For the original circuit, the frequency of observing the initial state is between $0$ and $7\%$. In other words, we have similar statistics for both the member and non-member words. However, for the optimized circuit, the initial state is observed visibly higher ($15\%$) for the input of length 37, which is a member word. Compared to the previous work on QFA implementation on IBM machines, we observed such improvement for the first time.

\begin{figure}[h!]
\begin{subfigure}{0.5\textwidth}
\centering
\includegraphics[width=5cm]{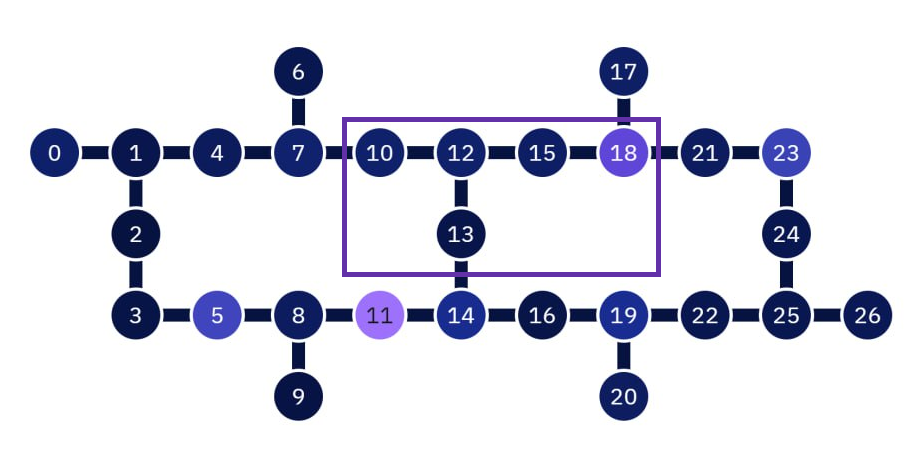}
\caption{5 working qubits  selected by the transpiler for the original circuit implementation.}
\label{fig:cairoOriginal}
\end{subfigure}
\begin{subfigure}{0.5\textwidth}
\centering
\includegraphics[width=5cm]{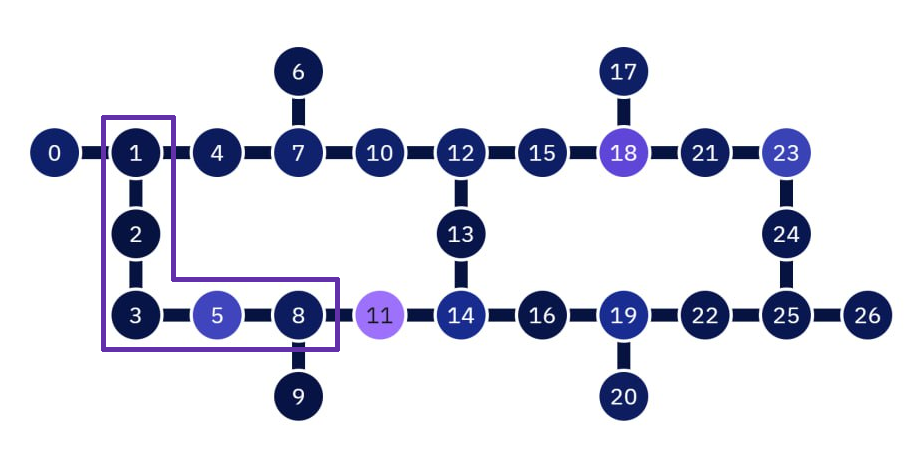}
\caption{5 working qubits  selected by the transpiler for the proposed circuit implementation.}
\label{fig:cairoLNN}
\end{subfigure}
\caption{$\mathtt{ibmq}\_\mathtt{cairo}$ machine qubits topology.}
\end{figure}

Our proposed circuit uses LNN architecture of qubits. So, the transpiler selects 5 linearly connected qubits (see bordered qubits in Fig.\ref{fig:cairoLNN}). While, for the original circuit, the transpiler selects more successfully structured qubits (see Fig.\ref{fig:cairoOriginal}). Despite that, we obtained better results for our proposed circuit.

\section{Conclusion}
\label{sec:con}
\vspace{-0.3cm}
For the current NISQ devices, noise poses a significant challenge both for quantum algorithm developers and physicists, as it can impact the accuracy of the implementation results. In this work, we presented different methods to improve the implementation of the QFA algorithm recognizing the language $\modp$ when implemented on real quantum devices. The methods focus on reducing the computational resources, in particular, the number of CNOT gates and depth. 

Furthermore, we considered devices with linear nearest neighbour topology and proposed an optimized circuit for the pseudo implementation of the QFA. We executed the circuit on real devices and observed that the number of CNOT gates for the proposed circuit after transpilation is approximately half of the transpilled original circuit. For $\mathtt{MOD_{37}}$ language, the new approach enables us to save up to 600 CNOT gates when processing the input string $a^{57}$. This reduction in the gate count significantly improves the accuracy of the results, as the original circuit with 1126 CNOT gates had a higher probability of computational errors, making it challenging to determine the validity of the results. Moreover, the experimental results showed the advantage of the proposed circuit implementation. Namely, it allows us to distinguish between members and non-members of the language, when the original circuit implementation  cannot do it.

\textbf{Acknowledgments} 
We sincerely thank our colleagues Kamil Khadiev, Mansur Ziatdinov, Aleksander Vasiliev, and Aida Gainutdinova for useful discussions. Part of this work was done by Khadevia during QCourse570-1 ``Projects in Quantum'' in Spring 2022 conducted by QWorld \& University of Latvia and supported by Unitary Fund. The research in Section 3 has been supported by the Kazan Federal University Strategic Academic Leadership Program ("PRIORITY-2030").
The research in Sections 4 and 5 is supported by Russian Science Foundation Grant 24-21-00406, https://rscf.ru/en/project/24-21-00406/. 

Salehi was partially supported by Polish National Science Center under the grant agreement 2019/33/B/ST6/02011. 

Yakary{\i}lmaz was partially supported by the Latvian Quantum Initiative under European Union Recovery and Resilience Facility project no. 2.3.1.1.i.0/1 /22/I/CFLA/001, the ERDF project Nr. 1.1.1.5/19/A/005 ``Quantum computers with constant memory'', and  the ERDF project number 1.1.1.5/18/A/020 ``Quantum algorithms: from complexity theory to experiment''.

\bibliographystyle{splncs04}
\bibliography{_ref}

\end{document}